\documentclass [preprint, prd, aps, nofootinbib]{revtex4}

\begin{document}
\draft
\title{\large \bf Neutrix Calculus and Finite Quantum Field Theory}
\author{\bf Y. Jack Ng\footnote{E-mail: yjng@physics.unc.edu} and H. van Dam}
\address{Institute of Field Physics, Department of Physics and
Astronomy,\\
University of North Carolina, Chapel Hill, NC 27599-3255\\}

\begin{abstract}
In general, quantum field theories (QFT) require regularizations and 
{\it infinite}
renormalizations due to ultraviolet divergences in their loop calculations.
Furthermore, perturbation series in theories like QED are not convergent
series, but are asymptotic series.  We
apply neutrix calculus, developed 
in connection with 
asymptotic series and divergent integrals, to QFT, obtaining 
{\it finite} renormalizations.
While none of the physically
measurable results in renormalizable QFT is changed,
quantum gravity is rendered more manageable  
in the neutrix framework. 

\bigskip


\end{abstract}
\maketitle
\newpage

The procedure of regularization
and renormalization is a big step
forward in making sense of the infinities that one encounters in
calculating perturbative series in quantum field theories. 
The result is a phenomenal success.  For example, 
Quantum Electrodynamics (QED), the paradigm of relativistic quantum field
theories, suitably regularized and renormalized, is arguably the most accurate
theory ever devised by mankind.  Yet in spite of the impressive
phenomenological successes, the specter of infinite renormalizations has
convinced many, including such eminent physicists as Dirac and Schwinger,
that we should seek a better mathematical and/or physical foundation 
for quantum field theory, without simultaneously tearing down 
the towering edifice we have built on the existing one.  
In another development, Dyson \cite{Dyson} has
shown that the series as defined by the Feynman rules in QED is not a
convergent series and has suggested that it is instead
an asymptotic series in the fine structure constant
$\alpha$, i.e., in the number of internal integrals (for given outside
lines).  In this paper, we propose to apply neutrix calculus, 
in conjunction with Hadamard integrals,
developed by J.G. van der Corput \cite{vdC}
in connection with asymptotic series and divergent integrals, 
to quantum field theories in general, and QED in particular, to
obtain finite results for the coefficients in the perturbation series.
(A more detailed discussion\cite{ngvd} will appear elsewhere.)
The replacement of regular integrals by Hadamard integrals in 
quantum field theory appears to make good mathematical sense, as 
van der Corput observed that Hadamard integrals are the proper tool to 
calculate the coefficients of an asymptotic series.
(Actually Hadamard integrals work equally well for convergent series.) 

We begin by recalling the definition of asymptotic series\cite{bender}.  The 
series $f(x) = a_0 + a_1 (x - b) + a_2 (x-b)^2 + ...$ for finite $b$ is an 
asymptotic series if and only if there exists an $n_0 > 0$, such that 
for $n > n_0$,
\begin{equation}
\lim_{x \rightarrow b} \frac{1}{(x-b)^n} \left| f(x) - a_0 - a_1(x-b)
- \dots - a_n (x-b)^n \right| = 0,
\label{asym}
\end{equation}
with the remnant being at most $\sim (x-b)^{n+1}$.

Next, following van der Corput \cite{vdC}, we define a neutrix
as a class of negligible functions defined in a domain, which satisfy the
following two conditions: (1) the neutrix is an additive group; (2) it does
not contain any constant except $0$.  Let us illustrate the concept with 
the following example considered by Hadamard: for $s$ real,
\begin{equation}
\int_{\xi}^2 x^{s-1} \, dx =
\left\{
\begin{array}{ll}
s^{-1} 2^s - s^{-1} \xi^s & \textrm{for}~ s \neq 0 \\
\log 2 - \log \xi & \textrm{for}~ s=0
\end{array}
\right.
\label{ex1}
\end{equation}
For $s > 0$, the integral converges even as $\xi \rightarrow 0$.  For
$s \leq 0$, Hadamard neglects $\xi^s / s$ and $\log \xi$ as $\xi 
\rightarrow 0$.  Here we have a neutrix which we will call $N(0)$, 
consisting of functions 
$\nu (\xi)= \epsilon (\xi) + c_1 \xi^s + c_2 \log \xi$, where 
$\epsilon(\xi) \rightarrow 0$ as $\xi \rightarrow 0$, and where 
$c_1$, $c_2$, and $s$ are arbitrary constants.  
This results in writing
\begin{equation} 
\int_{N(0)}^2 x^{s-1} \, dx = 
\left\{
\begin{array}{ll}  
s^{-1} 2^s & \textrm{for}~ s \neq 0\\
\log 2 & \textrm{for}~ s = 0 
\end{array} 
\right.
\label{ex11}
\end{equation}
Note the analytic extension in the complex $s$ plane of the answer
for Re $s > 0$ to the entire complex plane with the exclusion of 
$s=0$. 

Before applying neutrices to QED, we need to consider the generalized
Hadamard neutrix $H_a$ defined to contain the negligible functions
\begin{equation}
\nu(\xi) = U(\xi) + \epsilon (\xi),
\label{H(a)}
\end{equation}
where $\epsilon(\xi) \rightarrow 0$ as $\xi \rightarrow a$.  Each of the
functions $U(\xi)$ is defined by an asymptotic series based on $a$:
\begin{equation}
U(\xi) \sim \Sigma_{h=0}^{\infty} \chi_h \, (\xi - a)^{\Psi_h} \, 
\log^{k_h} (\xi-a).
\label{U(x)}
\end{equation}
Here $\chi_h$, $\Psi_h$ and integers $k_h \geq 0$ are independent
of $\xi$, Re $\Psi_h \rightarrow \infty$ as $h \rightarrow
\infty$, and $\log^{k_h} (\xi)$ stands for $(\log (\xi))^{k_h}$.
An example is provided by
\begin{equation}
\int_{H_a}^b (z-a)^{-1} \, \log^k (z-a) \, dz = (k+1)^{-1} \log ^{k+1} (b-a).
\label{exHa}
\end{equation}
Similarly, we can define the Hadamard neutrix $H_\infty$ by Eq.~(\ref{H(a)})
where now $\epsilon(\xi) \rightarrow 0$ as $\xi \rightarrow \infty$ and
the function $U(\xi)$ has a Hadamard development in powers of $\xi^{-1}$ in
its asymptotic series:
\begin{equation}
U(\xi) \sim \Sigma_{h=0}^{\infty} \chi_h \, \xi^{\Psi_h} \, 
\log^{k_h} \xi,
\label{U1(x)}
\end{equation}
where  Re $\Psi_h \rightarrow -\infty$ as $h \rightarrow
\infty$.

We can now demonstrate a very valuable property of the Hadamard neutrix.
Recall that in the theory of distributions developed by Schwartz, 
generalized functions usually cannot be multiplied.  Consider, for example,
the one-dimensional Dirac delta function multiplying itself 
$\delta (x) \times \delta (x)$.  This product is not mathematically 
meaningful because its Fourier transform diverges:
\begin{equation}
\int_{-\infty}^{\infty} {dk \over 2 \pi} \, 1 \times 1 \, 
\longrightarrow \infty,
\label{delsq}
\end{equation}
where we have used the convolution rule in Fourier transform 
and have noted that the
Fourier transform of the Dirac delta function is $1$.  In contrast, the
Hadamard method does allow multiplication for a wide class of 
distributions.  For the example of $\delta (x) \times \delta (x)$, the 
Hadamard-neutralized Fourier transform of the product
\begin{equation}
\int_{H_{-\infty}}^{H_{\infty}} {dk \over 2 \pi} \, 1 \times 1 \, = 0,
\label{delsq2}
\end{equation}
yields
\begin{equation}
\delta (x) \times \delta (x) = 0,
\label{delsq3}
\end{equation}
a mathematically meaningful (though somewhat counter-intuitive) result!

In doing quantum field theory in configuration space, we multiply
operator-valued distributions of quantum fields.  Or, in a slightly
different interpretation, we multiply  
singular functions such as Feynman propagators.  As we will see, the use of
neutrix calculus allows one to put these products on a mathematically sound   
basis.  Let us now generalize 
the above discussion for 1-dimensional Dirac delta functions to the 
case of 
(3 + 1)-dimensional Feynman propagators
\begin{eqnarray}
\Delta_{+} (x) &=& \int {d^4p \over (2 \pi)^4} {e^{ip \cdot x} \over 
{p^2 + m^2 -i\epsilon}}\\
&=& {1 \over 4 \pi} \delta (x^2) - {m \over 8 \pi  \sqrt{-x^2 - i\epsilon}}
H_1^{(2)} \! (m \sqrt{-x^2 - i\epsilon}),
\label{propag}
\end{eqnarray}
where $H^{(2)}$ is the Hankel function of the second kind
and we use the (+++-) metric.  The Fourier
transform of $\Delta_{+} (x) \times \Delta_{+} (x)$ (which appears in 
certain quantum loop calculations) is given by
\begin{eqnarray}
\int d^4x \, e^{-ip \cdot x} \, \Delta_{+} (x) \, \Delta_{+} (x)
&=& \int {d^4k \over (2 \pi)^4} \frac{1}{k^2 + m^2 - i\epsilon}
\frac{1}{(p - k)^2 + m^2 - i\epsilon}\\
&=& \frac{i}{4 (2 \pi)^2} D - \frac{i}{4 (2 \pi)^2}
\int_0^1 dz \, \log \left (1 + {p^2 \over m^2} z (1 - z) \right),
\label{Delsq}
\end{eqnarray}
where
\begin{eqnarray}
D &=& {1 \over i \pi^2} \int {d^4k \over (k^2 + m^2 )^2}\nonumber\\
&=& \int_0^\infty {k^2 dk^2 \over (k^2 +m^2)^2},
\label{sigma4}
\end{eqnarray}
with the second expression of $D$  
obtained after a Wick rotation.
But $D$ is logarithmically divergent.  Hence $\Delta_{+} (x) \times
\Delta_{+} (x)$ is not mathematically well defined.  Let us now see
how the Hadamard-van der Corput method gives mathematical meaning to 
this product.  Obviously, it 
is in the calculation of the logarithmically divergent $D$ 
where we apply neutrix calculus.
Introducing dimensionless variable $q = k^2/m^2$, we
bring in $H_\infty$ to write $D$ as
\begin{equation}
D = \int_0^{H_{\infty}} {q dq \over (q + 1)^2} = -1,
\label{D}
\end{equation}
where we have recalled that, for $q \rightarrow \infty$, $\log q$ is 
negligible in the Hadamard neutrix $H_\infty$.  It follows that, in 
the neutralized version, $\Delta_{+} (x) \times \Delta_{+} (x) \sim
\delta^{(4)} (x) +$ regular part
(where $\delta^{(4)} (x)$ is the 4-dimensional Dirac delta function), 
a much more mathematically palatable object.

As the first example in the application of neutrices to QED, let us 
consider the one-loop contribution to the electron's self energy
\begin{equation}
\Sigma (p) = -ie^2 \int {d^4k \over (2 \pi)^4} \frac {\gamma_{\mu}
 [- \gamma \cdot (p-k)+m] \gamma^{\mu}}{[k^2 + \lambda^2][(p-k)^2 + m^2]},
\label{sigma}
\end{equation}
where $m$ is the electron bare mass and we have given the photon a fictitious 
mass $\lambda$ to regularize infrared divergences. Expanding $\Sigma(p)$
about $\gamma \cdot p = -m$,
\begin{equation}
\Sigma (p) = \mathcal{A} + \mathcal{B} (\gamma \cdot p + m) + \mathcal{R},
\label{sigma1}
\end{equation}
one finds (cf. results found in Ref.\cite{JR})
\begin{equation}
\mathcal{A} = -{ \alpha \over 2 \pi} m \left({3 \over 2} D + {9 \over 4}
\right),
\label{sigma2}
\end{equation}
\begin{equation}
\mathcal{B} = - {\alpha \over 4 \pi} \left( D - 4 \int_{\lambda \over m}
^1 {dx \over x} + {11 \over 2} \right),
\label{sigma3}
\end{equation}
where $\alpha = e^2/4 \pi$ is the fine structure constant and
$D$ is given by Eq.~(\ref{sigma4}) and Eq.~(\ref{D}) 
in the pre-neutralized and neutralized forms respectively.
We note that $\mathcal{R}$, the last piece of $\Sigma (p)$ in 
Eq.~(\ref{sigma1}), is finite.
Mass renormalization and wavefunction renormalization are given by
$m_{ren} = m - \mathcal{A}$ and $\psi_{ren} = Z_2^{-1/2} \psi$ respectively
with $Z_2^{-1} = 1 - \mathcal{B}$.
Now, since $D = -1$ is finite,
it is abundantly clear that the renormalizations are {\it finite} in the 
framework of neutrix calculus.  There is no need for
a separate discussion of the electron vertex function renormalization
constant $Z_1$ due to the Ward identity $Z_1 =Z_2$.

The one-loop contribution to vacuum polarization is given by
\begin{equation}
\Pi_{\mu \nu} (k) = i e^2 \int {d^4p \over (2 \pi)^4} Tr \left (
\gamma_{\mu} {1 \over \gamma \cdot (p + \frac{k}{2}) + m}
\gamma_{\nu} {1 \over \gamma \cdot (p - \frac{k}{2}) + m} \right ).
\label{pi}
\end{equation}
A standard calculation\cite{JR} shows that $\Pi_{\mu \nu}$ takes on the form
\begin{equation}
\Pi_{\mu \nu} = \delta m^2 \eta_{\mu \nu} + (k^2 \eta_{\mu \nu} - k_{\mu} 
k_{\nu}) \Pi(k^2),
\label{pi1}
\end{equation}
where $\eta_{\mu \nu}$ is the flat metric (+++-),
\begin{equation}
\delta m^2 = {\alpha \over 2 \pi} (m^2 D + D'),
\label{pi2}
\end{equation}
and 
\begin{equation}
\Pi(k^2) = - {\alpha \over 3 \pi} (D + \frac{5}{6}) +
{2 \alpha \over \pi} \int_0^1 dx \, x (1-x) \, 
\log \left (1 + {k^2 \over m^2} x(1-x) \right ),
\label{pi3}
\end{equation}
with
\begin{equation}
D' = {1 \over i \pi^2} \int \frac {d^4p}{p^2 + m^2},
\label{pi4}
\end{equation} 
and $D$ given by Eq.~(\ref{sigma4}).  Just as $D$ is rendered finite upon 
invoking neutrix calculus (see Eq.~(\ref{D})), so is $D'$:
\begin{equation}
D' = m^2 \int_0^{H_{\infty}} {q dq \over q + 1} = 0,
\label{pi5}
\end{equation}
since both $q$ and $\log q$, for $q \rightarrow \infty$, are 
negligible in $H_\infty$.
Thus neutrix calculus yields a finite renormalization for both the photon 
mass and the photon wavefunction $A_{ren}^{\mu} = Z_3 ^{-1/2} A^{\mu}$
(and consequently also for charge $e_{ren} = Z_3^{1/2} e$) where $Z_3^{-1} = 1 
- \Pi(0)$. 
In electron-electron scattering by the 
exchange of a photon with energy-momentum $k$, vacuum polarization effects 
effectively replace $e^2$ by $e^2/(1 - \Pi(k^2))$, i.e.,
\begin{eqnarray}
e^2 \rightarrow e_{eff}^2 &=& \frac {e^2}{1 - \Pi (k^2)}\nonumber\\
&=& \frac {e_{ren}^2}{Z_3 (1 - \Pi (k^2))}\nonumber\\
&=& \frac {e_{ren}^2} {1 - (\Pi(k^2) - \Pi(0))}.
\label{run1}
\end{eqnarray}
Eq.~(\ref{pi3}) can be used, for $k^2 \gg m^2$, to show that
\begin{equation}
\alpha_{eff} (k^2) = \frac {\alpha}{1 - {\alpha \over 3 \pi} \log \left (
{k^2 \over \exp(5/3)\, m^2} \right )}.
\label{run2}
\end{equation}
Thus we have obtained the correct running of the coupling\cite{PS}
with energy-momentum in 
the framework of neutrices.  In fact, the {\it only} effect of neutrix calculus, 
when applied to QED (and other renormalizable theories), is to convert 
{\it infinite} renormalizations (obtained without using neutrix calculus) to 
mathematically well-defined {\it finite} renormalizations.  
As far as we can tell, {\it all} (finite)
physically observable results of QED are recovered. 
In passing we mention that the use of 
neutralized integrals does not affect the results of axial triangle anomalies.

As shown by the appearance of photon mass in the above discussion of 
vacuum polarization, the application of neutrix calculus to the 
energy-momentum cutoff regularization, though  
straightforward and natural, is ill suited for more complicated 
theories like those involving Yang-Mills fields.  For those theories, one 
should use other more convenient regularization schemes.
It is amusing to note that already in 1961 van der Corput suggested that,
instead of finding the appropriate neutrices, one can continue analytically
in any variable (presumably including the dimension of integrations) 
contained in the problem of tackling apparent divergences to calculate
the coefficients of the corresponding asymptotic series.  In hind sight, 
one recognizes 
that this was the approach taken by 't Hooft and Veltman who spearheaded
the use of dimensional regularizations\cite{PS}.   Let us now explore using 
neutrix calculus in conjunction with the dimensional regularization scheme.
In that case, negligible functions will include $1/\epsilon$ where
$\epsilon = 4 - n$ is the deviation of spacetime dimensions from 4.
In the one-loop calculations for QED, the 
internal energy-momentum integration is now over n 
dimensions.  The forms of $\Sigma (p)$ and $\Pi_{\mu \nu}$ remain the same as
given by Eqs.~(\ref{sigma1}) and (\ref{pi1}), but now with $\delta m^2 = 0$.  
Using the approximation for the gamma function,
$\Gamma(\epsilon) = \epsilon^{-1} - \gamma + \mathcal{O} (\epsilon)$, 
where $\gamma \simeq .577$ is the Euler-Mascheroni constant, 
and the approximation
%
$f^{\epsilon} \simeq 1 + \epsilon \log f$,
for $\epsilon \ll 1$,
one finds
\begin{eqnarray}
\mathcal{A} &=& \frac {\alpha m} {4 \pi} [3(\gamma - \log 4 \pi) +1]
+{\alpha \over 2 \pi} m \int_0^1 dx (1+x) \log D_0,\nonumber\\
\mathcal{B} &=& {\alpha \over 4 \pi}[1 + \gamma - \log 4 \pi]
+ {\alpha m^2 \over \pi}
\int_0^1 dx \frac{x(1 - x^2)}{m^2 x^2 + \lambda^2 (1-x)}
+ {\alpha \over 2 \pi}\int_0^1 dx (1-x) \log D_0,
\label{dim0}
\end{eqnarray}
where $D_0 = m^2x^2 + \lambda^2(1-x)$, and
\begin{eqnarray}
\Pi (k^2)      &=& {\alpha \over 3 \pi} [\gamma - \log 4 \pi]
+ {2 \alpha \over \pi}
\int_0^1 dx \, x(1-x)\, \log [m^2 + x(1-x)k^2],\nonumber\\
\frac {1}{Z_3} &=& 1 - {\alpha \over 3 \pi}[\gamma - \log 4 \pi]
- {\alpha \over 3 \pi} \log m^2.
\label{dim1}
\end{eqnarray}
%
By design, the generalized neutrix calculus renders all the renormalizations 
{\it finite}.  Again, {\it all} physically measurable results of QED appear 
to be recovered.
In this article we have explicitly considered QED to one-loop only.  But
we expect that
higher-loop calculations can be handled in the same way according to 
neutrix calculus.  It will be interesting to see explicitly
whether neutrix calculus, applied to higher-loop calculations, 
can provide new insights in the issue of overlapping divergences.

In the framework of quantum field theory for the four fundamental forces, 
the divergence problem is particularly severe for quantum gravity.  
Using dimensional regularization, 'tHooft and Veltman\cite{toast} found that
pure gravity is one-loop renormalizable, but in the presence of a scalar
field, renormalization was lost.  For the latter case, they found that the
counterterm evaluated on the mass shell is given by  
$\sim \epsilon^{-1} \sqrt{g} R^2$ with $R$ being the Ricci scalar.  Similar 
results for the cases of Maxwell fields and Dirac fields etc 
(supplementing the Einstien field) were obtained\cite{deser}.  It is 
natural to inquire whether the application of neutrix calculus could improve 
the situation.  The result is that now essentially the
divergent $\epsilon^{-1}$ factor is replaced by $-\gamma +$ constant.

It has not escaped our notice that neutrix calculus may ameliorate the 
hierarchy problem in particle physics. 
The hierachy problem is due to the fact that 
the Higgs scalar self-energies diverge quadratically, leading to a
stability problem in the standard model of particle physics.  But  
neutrix calculus treats quadratic divergences no different from  
logarithmic 
divergences, since both divergences belong to (the negligible functions of) 
the neutrix.
Neutrix calculus may also ameliorate the cosmological constant problem in 
quantum gravity.  
The cosmological constant problem can be traced to the quartic 
divergences in zero-point fluctuations from all quantum fields.  But again,
neutrix calculus
treats quartic divergences no different from logarithmic divergences.
Indeed, for a theory of gravitation with a cosmological constant term, 
the cosmological constant receives at most a finite renormalization from 
the quantum loops in the framework of neutrix calculus. 

We conclude with a comment on what neutrix calculus means to the general 
question of renormalizability of a theory.  We recall that a theory is
renormalizable if, in loop calculations, the counterterms vanish or if they
are proportional to terms in the original Lagrangian (the usual 
renormalization through rescaling).  It is still renormalizable if, to all
loops, the counterterms are of a new form, but only a finite number of such 
terms exist.  By this standard, neutrix calculus does not change the
renormalizability of a theory, since it merely changes potentially infinite 
renormalizations to finite renormalizations.  On the other hand, 
non-renormalizable terms, i.e., terms with positive superficial degree
of divergence, are tolerated in neutralized quantum field theory.
In a sense it is a pity that we have lost  
renormalizability as a physical restrictive criterion in the choice of 
sensible theories.  However, we believe that this is actually not as 
big a loss as it may first appear.  Quite likely, all realistic theories now 
in our possession are actually effective field 
theories.\cite{Schwinger,Weinberg}  They appear to be
renormalizable field theories because, at energies now accessible, or
more correctly, at sufficiently low energies, all the non-renormalizable 
interactions are highly suppressed.  By tolerating 
non-renormalizable terms, neutrix 
calculus has freed us from the past dogmatic and rigid requirement 
of renormalizability.  (Having said that, given a choice between 
renormalizable field theories and effective field theories, we 
still prefer the
former to the latter because of the former's compactness and predictive power.
But the point is that both types of theories can be accommodated in the 
framework of neutrix calculus.)  Furthermore, if the
application of neutrix calculus to loop calculations results in a 
term of a new form (like the Pauli term in QED) that is finite, then 
we have a prediction which, in principle, can be checked
against experiments to confirm or invalidate the theory in question.  For the
latter case, we will have to modify the theory by including a term of 
that form in the Lagrangian,
making the parameter associated with the new term an adjustable parameter 
rather than one that is predicted by the theory.  This loss of predictive
power is again not as big a loss as one may dread.

Lastly we should emphasize
that, for renormalizable theories as well as 
non-renormalizable theories (like quantum gravity?), neutrix calculus is
a useful tool to the extent that it is relevant for asymptotic series
and lessens the divergence of the theories.  
Based on our study so far, we tentatively conclude that
neutrix calculus has banished infinities from quantum field theory, rendering 
perturbative quantum field theory mathematically meaningful.

\bigskip

We thank J.J. Duistermaat, E. M. de Jager, T. Levelt, and
T. W. Ruijgrok for encouragement and for kindly providing us with relevant 
references
of the work by J.G. van der Corput.  We thank C. Bender and K. A. Milton for 
useful 
discussions.
We also thank L.~Ng and X.~Calmet for 
their help in the preparation of this manuscript.
We are grateful to the late Paul Dirac
and Julian Schwinger for inspiring us to look for a better mathematical
foundation for quantum field theory.  
This work was supported in part by DOE 
and by the Bahnson Fund of University of North
Carolina at Chapel Hill.

\end{document}